\documentclass[showpacs,preprintnumbers,amsmath,aps,amssymb]{revtex4}
\usepackage{graphicx}
\usepackage{dcolumn}
\usepackage{bm}
\begin{document}

\baselineskip 18pt

\title{Some exact solutions in  K-essence theory isotropic cosmology}

\author{Luis O. Pimentel$^1$}
\email{lopr@xanum.uam.mx}
\author{Carlos Gabarrete Fajardo$^2$ }
\email{carlos.gabarrete@gmail.com}
\affiliation{$^1$Departamento de F\'{\i}sica de la  Universidad Aut\'onoma Metropolitana\\
Apartado Postal 55-534, 09340, M\'exico, D.F.\\
$^2$ Departamento de Gravitaci\'on Altas Energ\'ias y Radiaciones, Escuela de F\'isica de la Universidad Nacional Aut\'onoma de Honduras}

\begin{abstract}
We use a simple form of the K-essence theory and apply it to the classic isotropic cosmological model and seek exact solutions.
The particular form of the kinetic term that we choose is  $K \left( 
\phi, X \right)= K_0(\phi)X^m +K_1$. The resulting field equations in the homogeneous and isotropic cosmology (FRW)is considered. Several exact solutions are obtained.  

\end{abstract}

  \maketitle                            

\section{Introduction}

In view of the problems with the $\Lambda CDM$ model, some alternative dynamical dark energy models have been advanced. Among them there is the K-essence theory that was developed in an attempt  to unify the description of dark matter, dark energy and inflation, by means of a scalar field  that has a non canonical kinetic term in the action\cite{s-b,armendariz,
bose, jorge,jorge1}.  This theory   is based on the idea of a dynamical attractor solution which  acts as a
cosmological constant only at the onset of matter domination and  then, K-essence dominates the matter density and produces a cosmic acceleration at around  the present epoch. The K-essence models are based in the following action that has a non canonical kinetic term for the scalar field
\cite{roland,chiba,bose,arroja,sn,tejeiro} .

\begin{equation}
\label{action}
S=  \int d^4 x \sqrt{-g} \left( \frac{R}{2\kappa^2} - K \left( 
\phi, X \right)+ L_{matter} \right)\, ,\quad X \equiv \partial^\mu \phi \partial_\mu \phi \, .
\end{equation}
 K-essence was
originally proposed as a model for inflation, and then as a model
for dark energy, along  as a  possibility to explore the  unification of  dark energy and
dark matter \cite{roland,bilic,bento}. Further motivations to
consider this type of theories come  from string theory
\cite{string}. For more details for the application of  K-essence to  dark energy
 can be  seen in reference \cite{copeland} and references therein. Some anisotropic 
 models have also  been considered before \cite{chimento,bianchi-ix}.

The form of the function $K \left( 
\phi, X \right)$ should be provided by a more fundamental theory like string theory or theories with more dimensions. Here we will consider the simple choice $K \left( 
\phi, X \right)= K_0(\phi)X^m +K_1$ in order to obtain exact solutions, where $ K_0(\phi)$ is a yet undetermined function and $K_1$ is a constant. In reference \cite{carlos} the particular choice $K_0(\phi)= K_0 {\phi}^n$ and $K_1=0$  were considered in order to obtain  explicit solutions for $\phi(t)$.

Variation of the action with respect to the metric gives the following equation:
\begin{equation}
\label{EE}
\frac{1}{\kappa^2}\left[ R_{\mu\nu} - \frac{1}{2}g_{\mu\nu} R \right] 
=  K \left( \phi, X \right) g_{\mu\nu} 
- 2 K_X \left( \phi, X \right) \partial_\mu \phi \partial_\nu \phi 
+ T_{\mu\nu}\, .
\end{equation}
Where $K_{X}(\phi,X)\equiv \partial K(\phi,X)/\partial X$, $T_{\mu\nu}$ is the energy momentum tensor of the matter content in the universe and we will take it  as a barotropic fluid. The variation with respect to the scalar field gives the following field equation: 
\begin{equation}
\label{scalar}
 - K_\phi \left( \phi, X \right) 
+ 2 \nabla^\mu \left( K_X \left( \phi, X \right) \partial_\mu \phi \right)\, = 0.
\end{equation}
Where $K_{\phi}(\phi,X)\equiv \partial K(\phi,X)/\partial \phi$. The metric of the isotropic and homogeneous  cosmology is chosen  in the following form:
\begin{equation}
	ds^2 = - dt^2 +a^2(t)\left[\frac{ dr^2}{1-k r^2} + r^2 d\theta^2+r^2 sin^2\theta d\varphi^2 \right].
	\label{metric}
\end{equation}

The equations that are obtained after the variation with respect to the metric and assuming a perfect fluid for the material content (using the fact that $\kappa^{2}=8\pi G$), are:
\begin{equation}
\frac{3 \dot{a}(t)^2}{a(t)^2}+\frac{3
   k}{a(t)^2}+8 \pi  G \left[K(\phi,X)+2
   {K_X(\phi,X)} \dot{\phi }(t)^2-\rho
   (t)\right]=0
   \label{eqe}
\end{equation}
\begin{equation}
\frac{2
  \ddot{a}(t)}{a(t)}+\frac{\dot{a}(t)^2}{a(
   t)^2}+\frac{k}{a(t)^2}+8 \pi  G
   \left[K(\phi,X)+P(t)\right]=0
   \label{eqf}
\end{equation}
Considering $K\left(\phi,X\right)= K_0(\phi)X^m +K_1$ and substituting in the equation field (3), we obtain:

\begin{equation}
\frac{6 m \dot{a}(t)}{a(t)}+\frac{(2
   m -1) {K_0}'(\phi
   (t))\dot{ \phi }(t)}{{K_0}(\phi
   (t))}+\frac{2m(2 m-1) \ddot{\phi
   }(t)}{\dot{\phi }(t)}=0.
   \label{scaleq}
\end{equation}
Here the dot means time derivative and the prime partial derivative with respect to $\phi$. The last equation (\ref{scaleq}) has the following first integral:
\begin{equation}
a  K_0(\phi)^{ \frac{2m-1 }{6 m}}\dot{\phi}^{\frac{2m-1 }{3}}=  c_1  ,
\label{firsti1}
\end{equation}
Here $c_1$ is the integration constant, that could be real or imaginary depending on the sign of $\phi$, $K_0$ and the value of $m$. When we obtain a solution for $a(t)$, we return to this first integral to obtain $\phi(t)$ by means of an integral (solution in quadrature).
For the material content we assume a barotropic fluid with equation of state $P=\epsilon \rho $. The conservation law for the fluid, then implies that $\rho= \rho_0 a^{-3(\epsilon +1)}$.   Substitution of the first integral (Eq.(\ref{firsti1})),$K \left( 
\phi, X \right)= K_0(\phi)X^m +K_1$)  into Eqs.(\ref{eqe} and \ref{eqf}) and noticing that for the metric under consideration $X= -{\dot{\phi }}^2$ we obtain 
\begin{equation}
 \dot{a}^2  +\left(\frac{8 {G\pi }}{3}\right)
   \left[ (-1)^m (1-2m) {c_1}^{\frac{-6 m}{1-2 m}}  a^{\frac{2(m+1)}{1-2
   m}}+{K_1} a^2 - \rho_0 a^{-(1+3\epsilon )} 
  \right]+ k=0,
\label{eqe1}
\end{equation}
\begin{equation}
\frac{2
  \ddot{ a}}{a}+\frac{\dot{a}^2
   }{a^2}+8 \pi  G \left[
   (-1)^m  {c_1}^{\frac{-6 m}{1-2 m}}  a^{\frac{6 m}{1-2
      m}} + K_1+ \epsilon \rho_0  a^{-3( \epsilon +1)} \right]
   +\frac{k}{a^2}=0.
\label{}
\end{equation}
Notice that Eq.(\ref{eqe1})is equivalent to the equation for the one dimensional motion of a particle under the action of the effective potential $V_{eff}$, 
\begin{equation}
V_{eff} = \left(\frac{8 {G\pi }}{3}\right)
   \left[ (-1)^m (1-2m) {c_1}^{\frac{-6 m}{1-2 m}} a^{\frac{2(m+1)}{1-2
   m}}+{K_1} a^2 - \rho_0 a^{-(1+3\epsilon )} 
  \right],
\label{veff}
\end{equation}
and with total  energy $E=-k$. Therefore we can  understand qualitatively the nature of the solutions, once $m$, $ \epsilon$  and $K_1$ are chosen, even when an analytic solution is not available. In the  next  sections we want to consider those cases in which  exact solutions in terms of ordinary functions are possible. 
\section{Only scalar field}
In this section we ignore the barotropic fluid and  consider the scalar field as the only component of the  material content of the cosmological model. The equation to solve is:
\begin{equation}
 \dot{a}^2  +\left(\frac{8 {G\pi }}{3}\right)
   \left[ (-1)^m (1-2m) {c_1}^{\frac{-6 m}{1-2 m}}  a^{\frac{2(m+1)}{1-2
   m}}+{K_1} a^2  
  \right]+ k=0,
\label{eqef}
\end{equation}

And we proceed to select some values of $m$ that allows us to have exact solution in terms of non special functions. 
\subsection{Case m=-1 }
Setting $m=-1$ in t Eq.(\ref{eqef}) the cosmological equation to solve is:
\begin{equation}
 \dot{a}^2  +\left(\frac{8 {G\pi }}{3}\right)
   \left[-3{c_1}^{2}+{K_1} a^2  
  \right]+ k=0,
\label{eqesc1a}
\end{equation}
We have the following solutions to Eq.(\ref{eqesc1a}) taking  different values for $K_1$, $k$ and $c_1$. When possible we take the initial condition $a(t)=0$.

For $ K_1>0$ and $8 \pi     G{c_1}^2 -k >0$
\begin{equation}
a(t)=\sqrt{\frac{3(8 \pi  G
   {c_1}^2 -k)}{8 \pi G K_1 }} \sin \left(
   \sqrt{\frac{8 \pi G K_1 }{3}}\; t
  \right).
\label{}
\end{equation}
This solution expands from a singularity to a maximum and then re collapses.

For  $ K_1=0$ and $8 \pi G
   {c_1}^2 -k >0$
\begin{equation}
a(t)=\sqrt{8 \pi G
   {c_1}^2 -k }\;t.
\end{equation}
Here we have a "coasting" solution starting from a singularity and expanding forever.

For  $ K_1<0$ and $8 \pi G {c_1}^2 -k >0$
\begin{equation}
a(t)= \sqrt{\frac{-3(8 \pi G {c_1}^2 -k )}{8 \pi G K_1}}sinh\left(\sqrt{\frac{-8 \pi G K1 }{3}}\; t\right).
\end{equation}
This solution starts from a singularity and expands without a re collapse.

For  $ K_1<0$  and  $8 \pi G {c_1}^2 -k <0$
\begin{equation}
a(t)=\sqrt{\frac{3(8 \pi G {c_1}^2 -k )}{8 \pi G K_1}}cosh\left(\sqrt{\frac{-8 \pi G K_1 }{3}}\; t\right).
\label{}
\end{equation}
This last  solution does not have a singularity, contracts from infinity to a minimum expansion factor and then expands forever.   
\subsection{Case m=0 }
When $m=0$ the corresponding cosmological equation is 
\begin{equation}
 \dot{a}^2  +\left(\frac{8 {G\pi }}{3}\right)
   \left[({K_1}+1) a^2  
  \right]+ k=0.
\label{eqm0}
\end{equation}
We can see that this equation is of the same form as Eq.(\ref{eqesc1a})but with different constant. The solutions will be as before but with the substitutions: $K_1 \rightarrow K_1+1$, $k \rightarrow k+8 \pi G c_1^{2}$. The solutions are as follows  

For $ K_1>-1$ and $ k =-1$
\begin{equation}
a(t)=\sqrt{\frac{3}{8 \pi G (K_1+1 )}} \sin \left(
   \sqrt{\frac{8 \pi G (K_1+1) }{3}}\; t
  \right).
\label{}
\end{equation}
For  $ K_1=-1$ and $k =-1$
\begin{equation}
a(t)=\;t
\end{equation}

For  $ K_1<-1$ and $k =-1$
\begin{equation}
a(t)= \sqrt{\frac{3}{8 \pi G (K_1+1 )}}\sinh \left(
   \sqrt{\frac{-8 \pi G (K_1+1) }{3}}\; t
  \right).
   \end{equation}

For  $ K_1<-1$ and $k =1$
\begin{equation}
a(t)= \sqrt{\frac{-3}{8 \pi G (K_1+1 )}}\cosh \left(
   \sqrt{\frac{-8 \pi G (K_1+1) }{3}}\; t
  \right).
   \end{equation}
%
\subsection{Case m=2 }
For $m=2$ the cosmological equation to solve is 
\begin{equation}
 \dot{a}^2  + 8\pi G \frac{K_1}{3} {a}^2 - \frac{8 \pi G c_1^2}{a^2}+k= 0,
\label{eqesc1c}
\end{equation}
In this case we have found solutions for $K1=0$

For $K_1=0$ and $k=0$
\begin{equation}
a(t)=\sqrt{\sqrt{32\pi G c_1^4}\; t}
  .
\label{}
\end{equation}   

For $K_1=0$ and $k\ne 0$
\begin{equation}
a(t)=\sqrt{\frac{8\pi G c_1^4}{k}- k\left(\; t- \sqrt{8\pi G c_1^4} \right)^2}  .
\label{}
\end{equation}  
In both cases $k=\pm 1$ the universe starts from a singularity but for $k=-1$ it expands forever and for $k=1$ there is a re collapse after The expansion factor reaches the maximum value of $a_M= \sqrt{8\pi G c_1^4}$. 
\section{Final remarks }
In this work we have considered the isotropic cosmology for some particular case of k-essence without extra matter content and have found several exact solutions. The  k-essence theory has been advanced as a possible model for dark energy and dark matter. More realistic models with dust and radiation might be  considered even if only numerical solutions are possible and the results will be reported elsewhere. 

\section{ Conflict of Interests}

The authors declare that there is no conflict of interests regarding the publication of this paper.

\acknowledgments
This work is part of the collaboration within the
Instituto Avanzado de Cosmolog\'{\i}a and Red PROMEP: Gravitation
and Mathematical Physics under project {\it Quantum aspects of
gravity in cosmological models, phenomenology and geometry of
space-time}. One of us (C.G.F.)thanks the MCTP for hospitality 
during the the elaboration of this work.

\end{document}